\documentclass[sigconf, preprint]{acmart}
\acmConference[arxiv]{}{April}{2024}

\renewcommand\footnotetextcopyrightpermission[1]{} 

\AtBeginDocument{%
  \providecommand\BibTeX{{%
    \normalfont B\kern-0.5em{\scshape i\kern-0.25em b}\kern-0.8em\TeX}}}

\usepackage{algorithmic}
\usepackage{graphicx}
\usepackage{textcomp}
\usepackage{xcolor}

\usepackage{subfiles} 
\usepackage{multirow}
\usepackage{listings}
\usepackage{subcaption}
\usepackage{color,xcolor}
\usepackage{color,xcolor,colortbl}
\usepackage{xspace}
\usepackage{enumitem}
\usepackage{graphicx}
\usepackage{xr}
\usepackage{url}

\usepackage[most]{tcolorbox} 
\usepackage{booktabs}
\usepackage{diagbox}
\usepackage{amsmath}
\usepackage{array}

\usepackage{balance}

\newcommand{\tool}{VulEval\xspace}

\def\BibTeX{{\rm B\kern-.05em{\sc i\kern-.025em b}\kern-.08em
    T\kern-.1667em\lower.7ex\hbox{E}\kern-.125emX}}

\begin{document}
\begin{sloppypar}

\title{VulEval: Towards Repository-Level Evaluation of Software Vulnerability Detection
}



\author{Xin-Cheng Wen}
\affiliation{%
  \institution{Harbin Institute of Technology}
  \city{Shenzhen}
  \country{China}}
\email{xiamenwxc@foxmail.com}

\author{Xinchen Wang}
\affiliation{%
  \institution{Harbin Institute of Technology,}
  \city{Shenzhen}
  \country{China}}
\email{200111115@stu.hit.edu.cn}

\author{Yujia Chen}
\affiliation{%
  \institution{Harbin Institute of Technology,}
  \city{Shenzhen}
  \country{China}}
\email{yujiachen@stu.hit.edu.cn}

\author{Ruida Hu}
\affiliation{%
  \institution{Harbin Institute of Technology,}
  \city{Shenzhen}
  \country{China}}
\email{200111107@stu.hit.edu.cn}

\author{David Lo}
\affiliation{%
  \institution{Singapore Management University}
  \country{Singapore}}
\email{davidlo@smu.edu.sg}

\author{Cuiyun Gao}
\affiliation{%
  \institution{Harbin Institute of Technology}
  \city{Shenzhen}
  \country{China}}
\authornote{Corresponding author.}
\email{gaocuiyun@hit.edu.cn}






\begin{abstract}

Deep Learning (DL)-based methods have proven to be effective for software vulnerability detection, with a potential for substantial productivity enhancements for detecting vulnerabilities. Current methods mainly focus on detecting single functions (i.e., intra-procedural vulnerabilities), ignoring the more complex inter-procedural vulnerability detection scenarios in practice.  For example, developers routinely engage with program analysis to detect vulnerabilities that span multiple functions within repositories. In addition, the widely-used benchmark datasets generally contain only intra-procedural vulnerabilities, leaving the assessment of inter-procedural vulnerability detection capabilities unexplored.

To mitigate the issues, we propose a repository-level evaluation system, named \textbf{\tool}, aiming at evaluating the detection performance of inter- and intra-procedural vulnerabilities simultaneously.  
Specifically, \tool consists of three interconnected evaluation tasks: 
\textbf{(1) Function-Level Vulnerability Detection}, aiming at detecting intra-procedural vulnerability given a code snippet;
\textbf{(2) Vulnerability-Related Dependency Prediction}, aiming at retrieving the most relevant dependencies from call graphs for providing developers with explanations about the vulnerabilities; and \textbf{(3) Repository-Level Vulnerability Detection}, aiming at detecting inter-procedural vulnerabilities by combining with the dependencies identified in the second task.
\tool also consists of a large-scale dataset, with a total of 4,196 CVE entries, 232,239 functions, and corresponding 4,699 repository-level source code in C/C++ programming languages.
By evaluating 19 vulnerability detection methods on the data split randomly and by time respectively, we observe that the repository-level vulnerability detection framework outperforms the corresponding function-level methods, with an increase of 1.51\% in F1 score and 2.63\% in MCC on average. It indicates that incorporating vulnerability-related dependencies facilitates vulnerability detection. Our experimental results also demonstrate that the performance of program-analysis- and prompt-based methods are not affected when splitting the data by time.
In addition, for the seven dependency retrieval methods studied, we find that lexical-based methods yield superior results than semantic-based methods for identifying vulnerability-related dependencies.
Our analysis highlights the current progress and 
future directions for software vulnerability detection.

\end{abstract}


\maketitle

\section{Introduction}
\label{sec:introduction}
Software vulnerabilities, mostly caused by insecure code, can be exploited to attack software systems, and further cause the security issues such as system crash, data leakage, and even critical infrastructure damage.
In the past ten years, the number of software vulnerabilities has increased more than five times, rising from 5,697 in 2013 to 29,065 in 2023~\cite{Statista1}. This increasing growth in both the quantity and type of software vulnerabilities has led to increasing economic losses~\cite{loss}. For example, Clop ransomware has successfully extorted more than \$500 million from various organizations~\cite{mimecast}. 
Therefore, it is necessary to develop effective technologies for software vulnerability detection.

The existing vulnerability detection methods can be categorized into four types: program analysis-based, supervised learning-based, fine-tuning-based, and prompt-based methods. 
The traditional program analysis-based vulnerability detection techniques, such as INFER~\cite{INFER} and CheckMarx~\cite{CHECKMARX}, rely on predefined rules to identify vulnerabilities. These approaches are labor-intensive and inefficient due to the diverse types of vulnerabilities and libraries.
Deep learning (DL)-based approaches have emerged as effective solutions, exhibiting notable success by mitigating the reliance on domain expertise and 
enhancing the ability to detect a variety of software vulnerabilities~\cite{cheng2022bug}.
Early DL-based approaches use the supervised learning-based methods, which leveraged Convolutional Neural Networks (CNNs)~\cite{wu2017vulnerability, DBLP:conf/icse/WuZD0X022/vulcnn}, Recurrent Neural Networks (RNNs)~\cite{sysevr, russell}, and Graph Neural Networks (GNNs)~\cite{devign, reveal} for learning the vulnerability representation. 

Nevertheless, the effectiveness of these supervised learning-based approaches is limited
by the 
scarcity of vulnerability data~\cite{DBLP:journals/corr/abs-2307-09163/reason}.
The emergence of pre-trained models like CodeBERT~\cite{DBLP:conf/emnlp/FengGTDFGS0LJZ20/codebert} and UniXcoder~\cite{DBLP:conf/acl/GuoLDW0022/unixcoder}, which are trained on large-scale open-source code repositories, has notably propelled this domain forward. These methods, equipped with extensive general programming knowledge, can be fine-tuned with vulnerability
datasets to greatly enhance the vulnerability detection performance, denoted as fine-tuning-based methods.
Nowadays, the prompt-based techniques
have utilized Large Language Models (LLMs), such as 
LLaMA~\cite{DBLP:journals/corr/abs-2302-13971/llama} and CodeLlama~\cite{DBLP:journals/corr/abs-2308-12950/codellama}, for vulnerability detection, 
marking a trend of inclination towards unsupervised methodologies in the domain.

\begin{figure*}[t]
	\centering
    \includegraphics[width=0.99\textwidth]{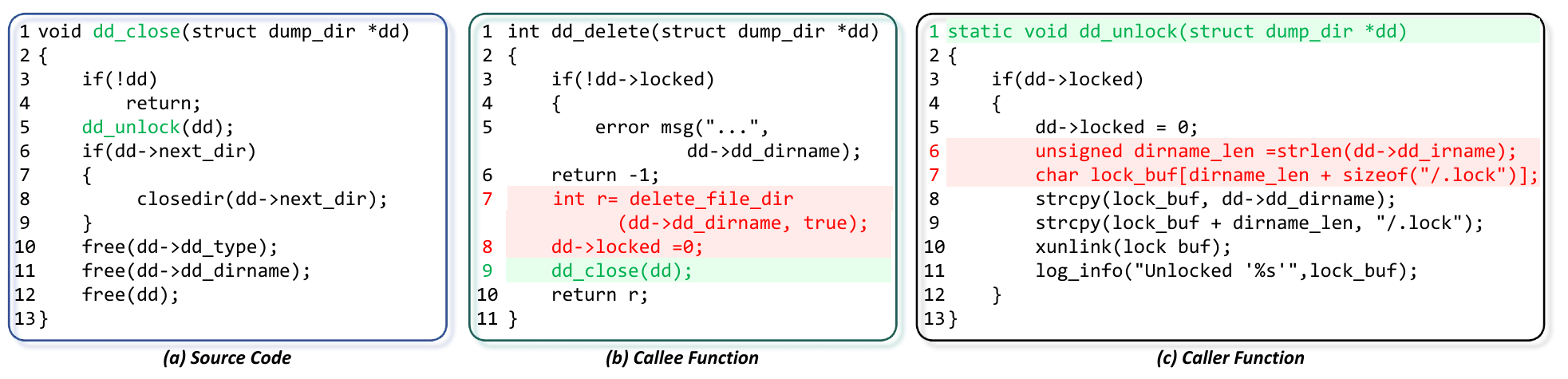}
	\caption{An inter-procedural vulnerability example of the CWE-20. Lines highlighted in green denote the call relation (i.e., callee and caller), and red denotes the vulnerable statements.}
	\label{codeexample}
\end{figure*}
Despite substantial
advancements in vulnerability detection through using
fine-tuning and prompt techniques,
evaluating the efficacy of these methods remains challenging.
Specifically, there exists a gap between the current evaluation
scenarios and real-world vulnerability detection scenarios, 
embodied in the following two aspects:

\textbf{(1) Lack of methods for detecting inter-procedural vulnerabilities.}
Despite the demonstrated efficacy of various methods for vulnerability detection, current evaluation frameworks primarily focus on the granularity of individual function or file, failing to fully account for the complexities of vulnerabilities that extend across multiple files or entire repositories. This narrow focus inadequately mirrors the complexity inherent in real-world vulnerability detection contexts, wherein developers routinely check with program analysis techniques to detect vulnerabilities that span multiple files in the repository level. 
For example, Figure~\ref{codeexample} presents an inter-procedural vulnerability of CWE-20 (Improper Input Validation)~\cite{CWE-20}. Figure~\ref{codeexample} (a), (b), and (c) illustrate the code snippet at the function level, the associated callee and caller functions, respectively. 
Specifically, the function \texttt{dd\_close} assumes that the \texttt{dd} pointer is non-null without verification, and proceeds to invoke \texttt{dd\_unlock} in Line 5 of Figure~\ref{codeexample}(a) and access member variables. It can cause the vulnerability (Lines 6-7 in Figure~\ref{codeexample}(c)). Similarly, \texttt{dd\_delete} performs an operation contingent upon the locked status without ensuring that the \texttt{dd} pointer is valid (Lines 7-8 in Figure~\ref{codeexample}(b)). 
Such inter-procedural vulnerabilities across multiple functions are hard to be identified by existing methods.

\textbf{(2) Lack of a comprehensive evaluation system for vulnerability detection.} 
The existing work generally conducts the evaluation on randomly split function-/file-level datasets, without considering different scenarios separately and the timeliness. 
The previous datasets~\cite{DBLP:conf/ndss/LiZXO0WDZ18/vuldeepecker, DBLP:conf/msr/FanL0N20/fan} only use the vulnerability patches to construct the dataset, which ignores the corresponding dependencies (e.g., callee and caller) in the repository.
In addition, due to the large number of dependencies from the call graph, it is necessary to retrieve
vulnerability-related dependencies for developers.
Furthermore, given the substantial vulnerabilities identified every year, the utilization of historical vulnerability data for detecting future vulnerabilities emerges as a critical need. However, the existing random-split setting may lead to risks of data leakage and the potential for inflated performance, which ultimately compromises the reliability of vulnerability detection methods and reflects the challenges present in real-world software development environments.


To mitigate the issues, in this paper, we propose a holistic evaluation system, named \textbf{\tool}, designed for evaluating inter- and intra-procedural vulnerabilities simultaneously. 
Specifically, we perform three interconnected tasks to construct the evaluation system:
\textbf{(1) Function-level Vulnerability Detection}, where the task is to predict the given code snippet whether it is vulnerable or not, aims at detecting intra-procedural vulnerability; \textbf{(2) Vulnerability-Related Dependency Prediction}, where the task is retrieving the 
vulnerability-related dependency from the call graph, thereby providing developers
with explanations about the vulnerabilities; and \textbf{(3) Repository-level Vulnerability Detection}, aiming at detecting inter-procedural vulnerabilities.
To explore the current vulnerability detection methods' performance in the third task, we propose a repository-level vulnerability detection framework
by combining dependencies identified in the second task.

We collect a large-scale
repository-level source code for each vulnerability patch to provide repository information.
It consists of 4,196 CVE entries, 232,239 functions, and corresponding 4,699 repository-level source code in C/C++ programming languages. We also extract 347,533 function dependencies (i.e., Callee and Caller) and 9,538 vulnerability-related dependencies from the repository to detect the inter-procedural vulnerability.

\begin{figure*}[t]
	\centering
 \includegraphics[width=0.9\textwidth]{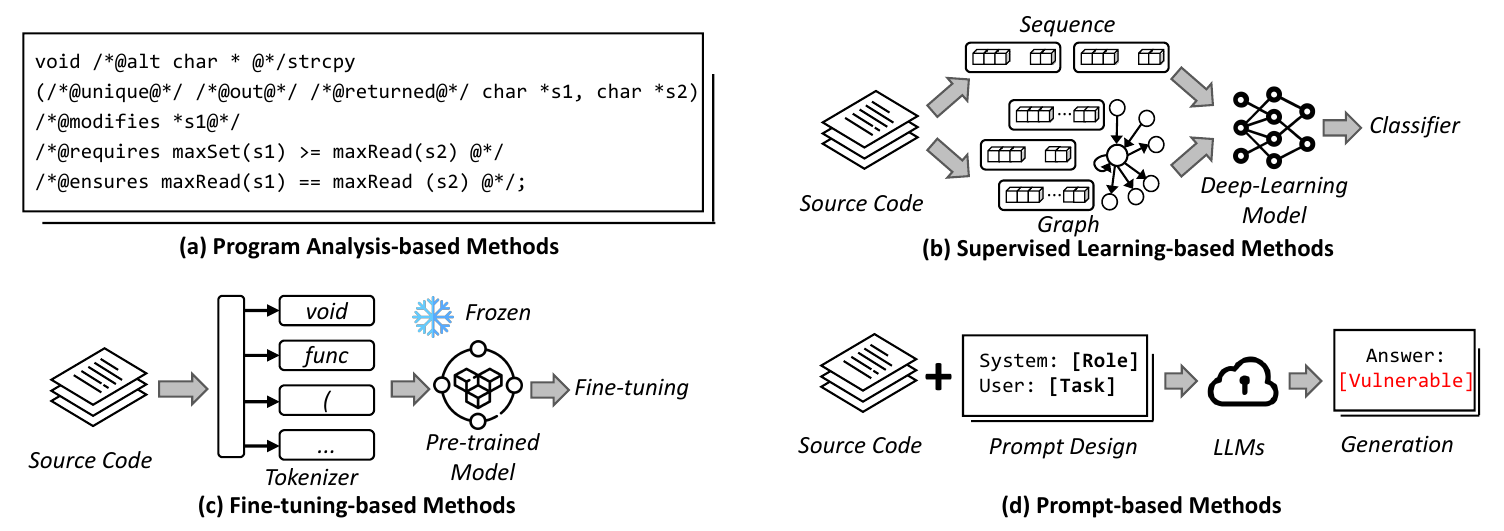}
	\caption{The four types of vulnerability detection methods.}
	\label{fig:background}
\end{figure*}

Based on the proposed evaluation system, we empirically study the performance of
the four types of vulnerability detection methods (i.e., 19 baselines) on \tool for function- and repository-level vulnerability detection. We also evaluate the three types of retrieval methods (i.e., seven
baselines) for vulnerability-related dependency prediction. During
the evaluation, we analyze the effectiveness in two settings (i.e., random split and split by time).
Furthermore, we highlight the current progress and shed light on future directions.

\textbf{Key Findings}. Based on the extensive experiments, our study reveals several key findings:
\begin{enumerate}
\item Incorporating contexts related to vulnerabilities in repository-level vulnerability detection enhances the performance compared with function-level vulnerability detection.
\item 
Supervised learning- and fine-tuning-based methods exhibit performance degradation within the time-split setting; while
the performance of program-analysis- and prompt-based methods are not affected. 
\item Lexical-based methods yield superior results than semantic-based methods in identifying vulnerability-related dependency. It is essential to develop more effective retrieval techniques for retrieving vulnerability-related dependencies. 


\end{enumerate}

\textbf{Contributions.} In summary, the major contributions of this paper are summarized as follows:
\begin{enumerate}
\item To the best of our knowledge, we are the first to propose a holistic evaluation system for evaluating inter and intra-procedural vulnerabilities simultaneously. 

\item We
collect a large-scale
repository-level source code and extract corresponding dependencies that provide repository-level information. 
We extract 347,533 dependencies and 9,538 vulnerability-related dependencies to detect the inter-procedural vulnerability.

\item We perform an extensive evaluation of 19 vulnerability detection methods and seven dependency retrieval methods in two settings. Our analysis highlights the current progress and future directions for software vulnerability detection.
\end{enumerate}


\section{Background}
\label{sec:background}

In this section, we introduce the existing vulnerability detection methods, including
program analysis-based, supervised learning-based, fine-tuning-based and prompt-based methods, as illustrated in Figure~\ref{fig:background}. 

\subsection{Program Analysis-based Methods}
Numerous program analysis-based methods have been proposed and widely used in the industry, such as CheckMarx~\cite{CHECKMARX}, FlawFinder~\cite{FLAWFINDER}, PCA~\cite{DBLP:conf/sigsoft/LiCSM20/pca} and RATs~\cite{RATs}. These methods leverage pre-defined rules or patterns
designed by experts to identify
specific types of vulnerabilities, such as stack-based buffer overflow, heap-based buffer overflow, and so on.

Figure~\ref{fig:background} (a) shows an 
example from
Splint~\cite{DBLP:journals/software/splint}. It represents a formal specification designed to express the expected behavior of the \texttt{strcpy} function and 
concurrently provides a rule for detecting potential buffer overflow vulnerabilities. 
Specifically, it 
checks the call complies with the condition \texttt{maxSet(s1) $>=$ maxRead(s2)}. If the Splint identifies any invocation that contravenes these conditions, it will alert the developer to a possible vulnerability. 
The advantage of these methods lies in their independence from extensive vulnerability datasets. Moreover, they explain the detected vulnerabilities by reporting the vulnerability-triggering path~\cite{cheng2022bug}. This path comprises a sequence of code snippets, thereby facilitating developers' verification processes.
However, designing
well-defined vulnerability rules or patterns is time-consuming and laborious~\cite{DBLP:journals/pacmpl/LiTMS18, DBLP:journals/toplas/LiTMS20}, making it challenging to cover all vulnerabilities.

\subsection{Supervised Learning-based Methods}
In recent years, many supervised-learning-based methods have been proposed that utilize representation learning techniques to capture vulnerability patterns. It mainly includes the sequence-based~\cite{DBLP:conf/ndss/LiZXO0WDZ18/vuldeepecker, sysevr} and graph-based~\cite{devign, reveal, IVDETECT} approaches. Figure~\ref{fig:background} (b) illustrates the process of these methods. The sequence-based methods typically use source code as input and learn the corresponding representations for determining whether the given code snippet is vulnerable or not. For instance, SySeVR~\cite{sysevr} extracts the code gadget and then uses the bidirectional Long Short-Term Memory network for vulnerability detection. VulCNN~\cite{DBLP:conf/icse/WuZD0X022/vulcnn} transforms source code into images and uses the CNNs to detect vulnerabilities.

Recent studies have shown that graph-based methods ascend in prominence due
to their superior interpretability and effectiveness. Compared to sequence-based approaches, these methods extract structured representations from source code, including Abstract Syntax Trees (AST), Control Flow Graphs (CFG), Data Flow Graphs (DFG), and Code Property Graphs (CPG)~\cite{DBLP:conf/icait/WangZWXH18/CPGVA}. Subsequently, GNNs are utilized to learn the graph representations for vulnerability detection.
In contrast to program analysis-based methods, these methods can
automatically capture vulnerability patterns, thereby mitigating the expenditure of human resources and time-consuming. Nevertheless, the effectiveness of these approaches highly depends on the availability of large and high-quality datasets for training. 

\subsection{Fine-tuning-based Methods}
Although supervised learning-based methods have demonstrated effectiveness for vulnerability detection, Croft et al.~\cite{DBLP:journals/corr/abs-2301-05456/dataquality} have pinpointed that existing vulnerability datasets often lack in quality and accuracy. It is challenging to apply them in real-world scenarios~\cite{DBLP:journals/corr/abs-2308-10523/PILOT}. 

Figure~\ref{sec:background} (c) shows the process of fine-tuning-based methods~\cite{DBLP:conf/iclr/GuoRLFT0ZDSFTDC21/graphcodebert, DBLP:conf/msr/FuT22/linevul,CodeGeeX}. These methods commence with pre-training on a vast corpus of code and textual data, and then
fine-tune the pre-trained model for a specific task.
For instance, CodeBERT~\cite{DBLP:conf/emnlp/FengGTDFGS0LJZ20/codebert} employs the Transformer architecture, utilizing an encoder for its training process. Similarly, CodeT5 \cite{DBLP:conf/emnlp/0034WJH21/CodeT5} and UniXcoder\cite{DBLP:conf/acl/GuoLDW0022/unixcoder} are specifically designed to provide both encoder and decoder in code-related tasks. Through the exploitation of knowledge encapsulated within pre-trained models, these approaches have been shown to excel in vulnerability detection. EPVD~\cite{DBLP:journals/tse/ZhangLHXL23/EPVD} introduces an algorithm for the selection of execution paths and leverages a pre-trained model to learn path representations. PILOT~\cite{DBLP:journals/corr/abs-2308-10523/PILOT} proposes a positive and unlabeled framework and uses the pre-trained model to construct the classifier.
However, these approaches are limited to
the length of input code and exhibit a deficiency in interpretability.

\subsection{Prompt-based Methods}
In recent years, LLMs have demonstrated superior performance in the fields of Software Engineering (SE)~\cite{DBLP:journals/corr/abs-2308-10620/se}
due to their broad generalization and reasoning abilities.
Prominent among these developments is the series of generative pre-trained transformer models, developed by OpenAI, including ChatGPT~\cite{ChatGPT} and GPT-4~\cite{GPT4}, as well as the LLaMA models unveiled by Meta, comprising both LLaMA~\cite{DBLP:journals/corr/abs-2302-13971/llama} and LLaMA2~\cite{DBLP:journals/corr/abs-2307-09288/llama2}. 
Figure~\ref{fig:background} (d) presents the process of prompt-based methods~\cite{DBLP:journals/corr/abs-2310-09810/llm4vul, ChatGPT}. It takes
source code as input, subsequently constructs a prompt tailored for
vulnerability detection, 
and feeds this prompt
into the LLMs. 
Then, the LLMs generate a response to detect whether the source code is vulnerable or not.
However, these LLMs face notable challenges in
software vulnerability detection~\cite{DBLP:journals/corr/abs-2310-09810/llm4vul}, which
primarily stems from two 
aspects. 
First, the code snippets often lack
enough contextual information 
for effectively detecting vulnerabilities.
Second, LLMs 
lack the specific domain knowledge required for vulnerability detection, which significantly hampers their performance.

\section{\tool System}
\label{sec:architecture}
\begin{figure*}[t]
	\centering
    \includegraphics[width=0.9\textwidth]{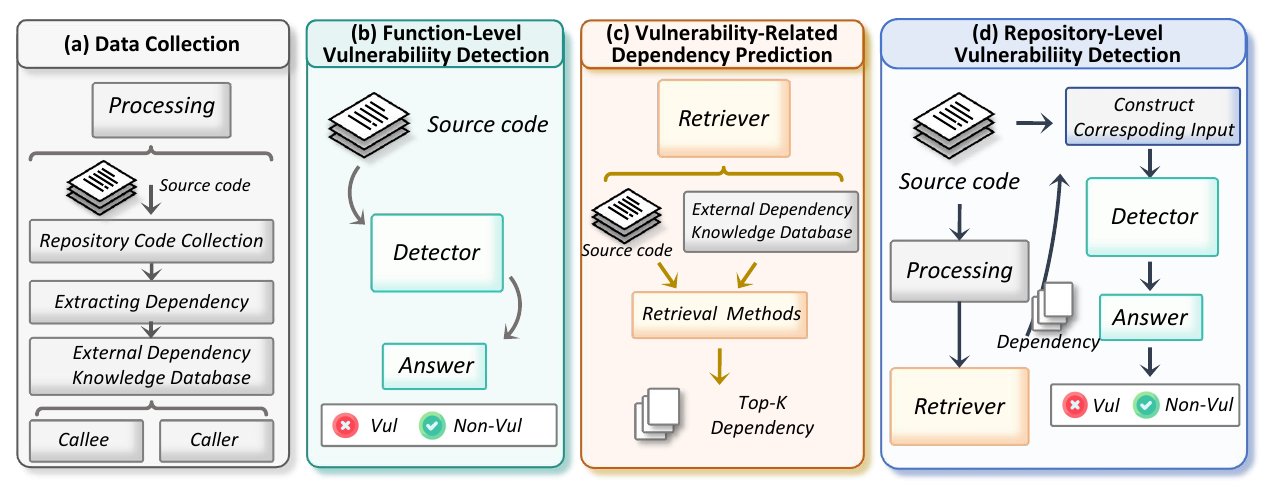}
	\caption{The overview
 of \tool. Figure (a), (b), (c), and (d) denote the process of data collection, function-level vulnerability detection, vulnerability-related dependency prediction, and repository-level vulnerability detection, respectively. }
	\label{architecture}
\end{figure*}

In this section, we introduce the evaluation system of \tool. It mainly includes two parts:
data collection, and evaluation task.

\subsection{Data Collection}

\subsubsection{Data Source}
Following the previous work~\cite{reef}, the raw data used to build \tool consists of a vast collection of CVE entries from the Mend~\cite{Mend}. The dataset consists of a total of 4,196 CVE entries, 4,699 vulnerability patches, and 164 vulnerability types in C/C++ programming languages. 

\subsubsection{Repository Code Collection}
For evaluating the inter-procedural vulnerabilities, we further collect the repository source code via three steps: 
(1) We select the repositories from which we can retrieve complete source code and commit logs via GitHub, Chrome, and Android.
(2) For each vulnerability patch, we
gather the repository-level source code corresponding to the commit time of the vulnerability patches. 
(3) For each file in the vulnerability patch, we use
Tree-sitter~\cite{Tree-sitter} to slice it as
function-level code snippets, where each function-level code snippet contains the corresponding repository-level source code separately. 

As shown in Table~\ref{dataset}, we collect
4,699 repository-level source code for vulnerability detection. In repository-level vulnerability detection, we also utilize the function-level label of the target function as the repository-level label (i.e., ``1'' for vulnerability and ``0'' for non-vulnerability). The target function and the corresponding dependencies are used as a whole sample to serve as the input for the repository-level sample.
\begin{table}[t]
\centering
\setlength{\tabcolsep}{2mm}
\renewcommand{\arraystretch}{1.1}

\caption{Statistics of the dataset.}
\resizebox{0.47\textwidth}{!}{
\begin{tabular}{ccccc}

\toprule
\rowcolor[HTML]{DEDEDE}
\textbf{Set} & \textbf{\# Function} &  \textbf{\# Repository}  & \textbf{\# Dependency} & \textbf{\# Vul-Dependency} \\

\midrule
Train  & 185,791/185,656   & 3,537/2,872 & 277,408/253,063   & 7,580/6,848   
\\
Valid  & 23,224/23,312   & 2,970/349  & 37,176/40,619   & 957/813          \\
Test & 23,224/23,271  & 2,984/331 & 32,949/53,851  & 1,001/1,877        \\
\midrule

All & 232,239 & 4,699 & 347,533 & 9,538 \\

\bottomrule

\end{tabular}
}
\label{dataset}
\end{table}


\subsubsection{Contextual Dependency Extraction}
One of the major contributions of \tool is that \tool considers the target code snippet's contextual dependency, which refers to the external code functions that are essential for vulnerability detection.

We extract the contextual dependencies of a code snippet through
program analysis of its belonging repository with two steps. 
(1) Before the extraction process, we first construct the repository database for each vulnerability patch, which includes the corresponding repository source code with
different header files (i.e., $.h$ ) and source code files (i.e., $.c$ and $.cpp$). (2) Then, we select the code changed file in the vulnerability patch and employ static program analysis tool~\cite{cflow} to extract the dependency elements. We classify them into the ``Callee'' and ``Caller'' dependencies. Specifically, ``Callee'' represents the user-defined function being invoked or executed by the vulnerability patch. The ``Caller'' denotes the user-defined function of the repository source code responsible for invocating the function in the vulnerability patch. 

As shown in Table~\ref{dataset}, we
extract
347,533 dependencies in the repository-level source code.
We also label 9,538 vulnerability-related dependencies (i.e., denoted as ``Vul-Dependency''), which are directly involved in code changes of vulnerability patches.
All the other dependencies are considered unrelated to the vulnerability.



\subsection{Evaluation Task}
\tool involves three evaluation tasks: function-level vulnerability detection, vulnerability-related dependency prediction, and repository-level vulnerability detection, with details as below.

\subsubsection{Function-level Vulnerability Detection (Detector)}
This task aims to predict whether the function contains a vulnerability or not. As shown in Figure~\ref{architecture}, Function vulnerability detection focuses solely on the source code of the target prediction function as input, abstaining from incorporating any inter-procedure information beyond the function itself. 
The goal of this task is to learn a detector $f$ that can be illustrated as follows:
\begin{equation}
\label{FVD}
f:\mathcal{X}\mapsto\mathcal{Y}, \mathcal{Y}=\{0,1\}
\end{equation}
where $\mathcal{X}$ denotes the input of function-level code snippet and $\mathcal{Y}$ denotes
the label which is set as
1 for vulnerable code snippets and 0 otherwise.
\subsubsection{Vulnerability-Related Dependency Prediction (Retriever)}
The task aims at
providing developers with explanations about the vulnerabilities.
Table~\ref{dataset} shows that the dataset has 347,533 dependencies, but only 9,538 dependencies are related to vulnerabilities. 
Thus, it is necessary to retrieve
vulnerability-related dependencies from the large number of dependencies in the repository source.
As shown in Figure~\ref{architecture} (c), the process of dependency prediction generally involves the ``Callee'' ($Callee$) and ``Caller'' ($Caller$) dependency extracted from the input function $\mathcal{X}$, followed by the calculation of the degree of vulnerability-related between the input code snippet and each candidate dependency. The general retrieval function $g$ for identifying vulnerability-related
dependency can be formulated as follows:
\begin{equation}
\label{RSD}
\mathop{max}_{i, j \in \{1,2,..,m+n\}}^{k} g(\mathcal{X},Callee^{i},Caller^{j})
\end{equation}
where $Callee^{i}$ and $Caller^{j}$ represent the $i$-th and $j$-th candidate dependency, respectively, and $m$ and $n$ are
the number of ``Callee'' and ``Caller'' candidate dependencies, respectively. $k$ denotes the top $k$ relevant dependencies to be retrieved in this task. 
\subsubsection{Repository-level Vulnerability Detection}
Repository-level vulnerability detection is our proposed
task, 
which integrates dependencies identified in the second task 
for vulnerability detection, as shown in Figure~\ref{architecture} (d). 
It first uses the ``Retriever'' to retrieve the associated dependency of the given code snippet. Then, the identified dependencies (i.e., retrieved by  ``Retriever'), are concatenated with the target function as
input.
Then, it uses the ``Detector'' to determine whether the input is vulnerable
or not. The definition
of repository-level vulnerability detection $h$ can be represented as follows:
\begin{equation}
\label{FVD}
h: (\mathcal{X}, Callee_\mathcal{X}, Caller_\mathcal{X})\mapsto\mathcal{Y}, \mathcal{Y}=\{0,1\}
\end{equation}
where the $Callee_\mathcal{X}$ and $Caller_\mathcal{X}$ denote the retrieved ``Callee'' and ``Caller'' dependency from code snippet $\mathcal{X}$, respectively.

\section{EXPERIMENTAL Setup}
\label{sec:evaluation}

\subsection{Research Questions}
Our experiment intends to answer the following research questions:
\begin{itemize}
\item \textbf{RQ1}: How do program analysis-, supervised learning-, fine-tuning- and prompt-based methods perform in function-level vulnerability detection?
\item \textbf{RQ2}: How do the retrieval methods perform in identifying the vulnerability-relevant dependency?
\item \textbf{RQ3}: How do these methods perform in repository-level vulnerability detection?
\item \textbf{RQ4}: How do these vulnerability detection methods perform for each CWE type?
\end{itemize}
\subsection{Experimental Methodology}

\subsubsection{Comparison on Vulnerability Detection Approaches.}
To evaluate the efficacy of vulnerability detection across function-level and repository-level contexts, our benchmark compares four types of vulnerability detection approaches: 
\textbf{(1) Program analysis-based methods}: Following the previous works~\cite{DBLP:conf/icse/WuZD0X022/vulcnn, DBLP:conf/icse/CaoSBWLT22/mvd}, we select four popular program analysis-based vulnerability detectors, i.e., Cppcheck~\cite{Cppcheck}, Flawfinder~\cite{FLAWFINDER}, RATS~\cite{RATs}, and Semgrep~\cite{Semgrep}. These methods leverage predefined rules and patterns to discern potentially improper operations within source code.
\textbf{(2) Supervised learning-based methods}: We use Devign~\cite{devign} and Reveal~\cite{reveal} as representative supervised baselines,
which are widely adopted as baselines in recent works~\cite{DBLP:journals/infsof/CaoSBWL21, DBLP:journals/corr/abs-2308-10523/PILOT, IVDETECT}. 
These methods construct graphs from source code and then perform vulnerability detection using features obtained through Gated Graph Neural Networks~\cite{ggnn}.
\textbf{(3) Fine-tuning-based methods}: The fine-tuning-based methods consists three general pre-trained models and four state-of-the-art approaches specialized for vulnerability detection. 
We select three general pre-trained models, CodeBERT~\cite{DBLP:conf/emnlp/FengGTDFGS0LJZ20/codebert}, CodeT5~\cite{DBLP:conf/emnlp/0034WJH21/CodeT5}, and UniXcoder~\cite{DBLP:conf/acl/GuoLDW0022/unixcoder}, for their widespread adoption in code-related tasks and further fine-tune these models for vulnerability detection. In addition, we choose four state-of-the-art models designed specifically for vulnerability detection, including PILOT~\cite{DBLP:journals/corr/abs-2308-10523/PILOT},
EPVD~\cite{DBLP:journals/tse/ZhangLHXL23/EPVD}, LineVul~\cite{DBLP:conf/msr/FuT22/linevul} and PDBERT~\cite{PDBERT}.
\textbf{(4) Prompt-based methods}: 
We choose two open-source LLMs: LLaMA~\cite{DBLP:journals/corr/abs-2302-13971/llama} and CodeLlama~\cite{DBLP:journals/corr/abs-2308-12950/codellama} for their proficiency in text and code generation, respectively. Additionally, we also incorporate two closed-source LLMs: ChatGPT (i.e., GPT-3.5-\textit{turbo}) and GPT-3.5-instruct, developed by OpenAI, 
which produce text with 175 billion parameters. 
For these LLMs, we utilize the process described in Section~\ref{sec:background} (d) to assess their effectiveness in vulnerability detection.

\subsubsection{Comparison on Dependency Prediction Approaches.}
We first extract all functions from the call graph as dependency candidates. Then, three types and seven baselines are employed for the vulnerability-related dependency prediction task: 
\textbf{(1) Random method:} 
This method retrieves code snippets randomly, 
serving as a foundational baseline for evaluating other prediction methods. To mitigate sampling bias, we repeat this randomized process 100 times and report the average results.
\textbf{(2) Lexical-based methods:} 
We evaluate the relevance of vulnerability dependencies using two primary metrics as baselines: Jaccard Similarity and Edit Similarity~\cite{DBLP:conf/sigsoft/SvyatkovskiyDFS20/jses}. Additionally, we use BM25~\cite{DBLP:journals/ftir/RobertsonZ09/BM25} and BM25+~\cite{DBLP:conf/adcs/TrotmanPB14/BM25plus} as lexical-based baselines weighting functions to rank dependencies by their relevance to specific code snippets. 
\textbf{(3) Semantic-based methods:} 
We leverage pre-trained models as backbone, specifically CodeBERT~\cite{DBLP:conf/emnlp/FengGTDFGS0LJZ20/codebert} and UniXcoder~\cite{DBLP:conf/acl/GuoLDW0022/unixcoder}) to obtain feature embeddings and then employ Cosine Similarity~\cite{DBLP:journals/isci/XiaZL15/cs} to measure the semantic relevance between code and dependency snippets.

\subsection{Evaluation Metrics}
\subsubsection{Metrics for Vulnerability Detection Task}
We use the following four widely-used performance metrics for vulnerability detection:

\textbf{Precision:} It is calculated as the ratio of true positives ($\text{TP}$) to the sum of true positives and false positives ($\text{FP}$), expressed as $\text{Precision} = \frac{\text{TP}}{\text{TP}+\text{FP}}$. It signifies the proportion of correctly identified vulnerabilities among all retrieved vulnerabilities.

\textbf{Recall:} Recall is computed as the ratio of $\text{TP}$ to the sum of $\text{TP}$ and false negatives ($\text{FN}$), given by $\text{Rec} = \frac{\text{TP}}{\text{TP}+\text{FN}}$. It represents the proportion of vulnerabilities detected by baselines out of all vulnerabilities.

\textbf{F1 Score (F1):} F1 score is defined as the harmonic mean of precision and recall, calculated using the formula $\text{F1} = 2 \times \frac{\text{Pre} \times \text{Rec}}{\text{Pre}+\text{Rec}}$. It serves as a combined measure of precision and recall, providing insight into the balance between them.

\textbf{Matthews Correlation Coefficient (MCC):} MCC is a measure of binary classification, particularly useful in imbalanced datasets, computed as $\text{MCC}=\frac{\text{TP} \times \text{TN} - \text{FP} \times \text{FN}}{\sqrt{(\text{TP} + \text{FP})(\text{TP} + \text{FN})(\text{TN} + \text{FP})(TN + \text{FP})}}$, where TN denotes the true negatives.

\subsubsection{Metrics for Dependency Prediction Task}
We propose the following two metrics for identifying dependency:

\textbf{Precision@K (Pre@K)}: It is the proportion of correctly predicted dependency amongst the Top-K predicted dependency, calculated as follows: 
$Pre@K = \frac{MATCH_k}{k}$. 
where $MATCH_k$ denotes the count of correctly predicted dependencies among the Top-K predicted dependency.

\textbf{Recall@K (Rec@K)}: It is the proportion of correctly predicted dependency amongst the ground-truth dependency, which is computed as $Rec@K = \frac{MATCH_k(m)}{GT}$, where GT represents the total count of ground-truth, vulnerability-related dependencies.

\subsection{Data Split}
In this paper, we experiment under the following two settings:
\textbf{(1) Random Split}: Following the previous work~\cite{devign, reveal}, we randomly split the datasets into disjoint training, validation, and test sets in a ratio of 8:1:1. 
\textbf{(2) Time Split}: 
To mitigate the risk of data leakage and effectively evaluate the methods' ability to identify emerging vulnerabilities, we adopt a time-split setting based on the ``commit date'' of vulnerability patches. We divide the dataset into training, validation, and test sets in an 8:1:1 ratio. Specifically, patches before 2018-03-21 are designated for training, those before 2022-07-21 constitute the validation set, and patches after this date are used for the test set.

\subsection{Implementation Details}
For program analysis-, supervised-, and fine-tuning-based methods, we directly use the replication packages and hyper-parameters that have been made publicly accessible.
For prompt-based methods, we downloaded LLaMA (i.e., 7B and 13B) and CodeLlama (i.e., 7B and 13B) from the HuggingFace Hub~\cite{HuggingFace} and deploy them locally by the vLLM~\cite{DBLP:conf/sosp/KwonLZ0ZY0ZS23/vllm} framework. For ChatGPT (``gpt-3.5-\textit{turbo-0301}'') and GPT-3.5-instruct (`gpt-3.5-\textit{turbo-instruct}''), we use the public APIs and initial parameters setting provided by OpenAI.
All evaluations are conducted on a server equipped with four NVIDIA A100-SXM4-40GB.

\section{Experimental Results}
\label{sec:experimental_result}
\begin{table*}[h]
\centering
\setlength{\tabcolsep}{2mm}
\renewcommand{\arraystretch}{1}
\caption{
The experimental results of function-level vulnerability detection in random and time split settings. 
Bold text cells represent the best performance. 
The cells in grey represent the performance of the top-3 best methods, with darker colors representing better performance.
}
\resizebox{0.90\textwidth}{!}{
\begin{tabular}{cl@{\hspace{2\tabcolsep}}cccc@{\hspace{2\tabcolsep}}cccc}
\toprule
\rowcolor[HTML]{DEDEDE}
\multicolumn{2}{c}{\textbf{Split Methods}}                                                          & \multicolumn{4}{c}{\textbf{Random}}                                                       & \multicolumn{4}{c}{\textbf{Time}}                                                         \\
[1pt] 
\rowcolor[HTML]{DEDEDE}
\textbf{Type}                & \textbf{Baseline}                                                    & \textbf{Precision}   & \textbf{Recall}      & \textbf{F1}    & \multicolumn{1}{l}{\textbf{MCC}}         & \textbf{Precision}   & \textbf{Recall}      & \textbf{F1}    & \textbf{MCC}         \\

\toprule
\multirow{4}{*}{Program Analysis}      & Cppcheck                                                             & 12.12                & 1.79                 & 3.12                 & 3.61               & \cellcolor{gray!40}19.43                & 4.34                 & 7.10                 & 7.45               \\
                             & Flawfinder                                                           & 6.54                 & 24.78                & 10.35                & 7.65               & 8.55                 & 32.52                & \cellcolor{gray!20}13.54                & 9.73               \\
                             & RATS                                                                 & 7.06                 & 12.54                & 9.04                 & 5.80               & 11.18                & 20.13                & \cellcolor{gray!70}\textbf{14.38}                & \cellcolor{gray!40}10.14               \\
                             & Semgrep                                                              & 10.36                & 7.76                 & 8.87                 & 6.64               & 8.37                 & 6.67                 & 7.42                 & 7.45               \\
\midrule

      & Devign                                                               & 38.36                & 24.26                & 29.72                &  28.88                    & 9.41                 & 5.26                 & 6.75                 &                3.95      \\

          \multirow{-2}{*}{Supervised Learning}                    & Reveal                                                               & 5.95                 & 33.35                & 10.08                &   7.96                   &     7.20                 &             24.68         &    10.99                  &              5.83        \\
\midrule
\multirow{7}{*}{\begin{tabular}[c]{@{}c@{}}Fine-tuning\end{tabular}} & CodeBERT                                                             & 51.45                & 31.79                & 39.30                & 39.09               & 13.85                & 2.86                 & 4.74                 & 4.56               \\
                             & CodeT5                                                               & \cellcolor{gray!20}51.83                & 35.97                & \cellcolor{gray!70}\textbf{42.47}                & \cellcolor{gray!70}\textbf{41.80}               & \cellcolor{gray!20}17.23                & 5.40                 & 8.23                 & 7.58               \\
                             & UniXcoder                                                            & \cellcolor{gray!70}\textbf{63.64}                & 18.81                & 29.03                & 33.66               & 13.36                & 4.13                 & 6.31                 & 5.31               \\
                             & PILOT                                                                 & 49.01                & 33.28                & 39.64                & 38.96               & 4.26                 & \cellcolor{gray!70}\textbf{91.63}                & 8.15                 & 2.74               \\
                             & EPVD                                                                 & 46.84                & 35.33                & 40.28                & 39.18               & 12.76                & 4.41                 & 6.55                 & 5.39               \\
                             & LineVul                                                              & 47.95                & 34.93                & \cellcolor{gray!20}40.41                & \cellcolor{gray!20}39.44               & 12.79                & 2.97                 & 4.82                 & 4.31               \\
                             & PDBERT                                                               & \cellcolor{gray!40}51.89                & 34.78                & \cellcolor{gray!40}41.64                & \cellcolor{gray!40}41.11               & \cellcolor{gray!70}\textbf{34.97}                & 5.30                 & 9.20                 & \cellcolor{gray!70}\textbf{12.32}               \\
\midrule

       & LLaMA-7B                                                             & 2.73                 & \cellcolor{gray!40}68.66                & 5.25                 & \multicolumn{1}{l}{-1.51}              & 4.17                 & \cellcolor{gray!20}70.66                & 7.88                 & 0.87               \\
                            & LLaMA-13B                                                            & 2.87                 & \cellcolor{gray!20}57.46                & 5.47                 & 0.00                 & 3.91        & 57.84       & 7.33        & -0.90       \\
                             & CodeLlama-7B                                                     & 0.88                 & \cellcolor{gray!70}\textbf{72.09}                & 1.74                 & {-2.94}              & 2.48                 & \cellcolor{gray!40}75.65                & 4.81                 & -3.33              \\
                            & CodeLlama-13B    &                                                2.26&
50.45&
4.33&
-4.98 & 3.28 & 53.60 & 6.18 & -5.51 \\
                             & GPT-3.5-instruct                                               & 4.02                 & 53.58                & 7.48                 & 5.37               & 5.10                 & 46.93                & 9.20                 & 4.09               \\
            \multirow{-7}{*}{\begin{tabular}[c]{@{}c@{}}Prompt\end{tabular}}                  & \begin{tabular}[c]{@{}c@{}}ChatGPT\end{tabular} & 7.38                 & 32.55                & 12.03                & \multicolumn{1}{l}{10.44}               & 9.69                 & 26.69                & \cellcolor{gray!40}14.22                & \cellcolor{gray!20}10.13               \\  
\bottomrule
\end{tabular}}
\label{RQ1}
\end{table*}
    \subsection{RQ1: Effectiveness in Function-level Vulnerability Detection}

\subsubsection{Effectiveness in Random Split Setting.}
To answer RQ1, we compare the four types of vulnerability detection methods, including program analysis-, supervised learning-, fine-tuning-, and prompt-based methods.
The results are shown in the middle
column of Table~\ref{RQ1}.

\textbf{Fine-tuning-based methods demonstrate superior performance compared to other methods in the random split.}
Specifically, these methods yield average results of 51.80\% in precision, 38.97\% in F1 score, and 39.03\% in MCC in the random setting. We also observe that fine-tuning-based methods fall short in terms of recall, with an average of 32.13\%, compared to prompt-based methods, which achieve an average result of 55.80\% in terms of recall. In the broader evaluation for Top-3 performance across four metrics (comprising 12 instances), these methods demonstrate superiority in 9 out of 12 cases,
achieving the highest precision, F1 and MCC, with a score at
63.64\%, 42.47\% and 41.80\%, respectively. 

\textbf{The program analysis- and supervised learning-based methods consistently exhibit inferior performance across all metrics.} 
Program analysis-based methods often target only specific vulnerability types and consequently yield poor results in general vulnerability detection. The pre-trained models utilized in fine-tuning and prompt-based methods learn more
general knowledge during the pre-training phase, thereby endowing them with superior efficacy than supervised learning-based methods. 

\subsubsection{Effectiveness in Time Split Setting.}

We also 
evaluate
all baseline methods in the time split setting
to comprehensively verify 
their
effectiveness 
against real-world scenarios without data leakage. The results are shown in the right column of Table~\ref{RQ1}. 

\textbf{Degradation in performance of supervised learning- and fine-tuning-based methods within time split.}
Analyzing the results in Table~\ref{RQ1}, we observe that 
fine-tuning-based methods exhibit a substantial decrease in all four metrics, showcasing an average decrement of 36.20\% in precision, 15.46\% in recall,	32.11\% in F1 score, and 33.00\% in MCC.
Similarly, the supervised learning-based methods also demonstrate a decline in performance across four metrics by 13.85\%, 13.84\%, 11.03\%, and 13.53\%, respectively. 
This can be attributed to their heavy reliance on 
extracting semantics from historical data, rather than directly capturing vulnerability patterns.
However, most vulnerabilities are discovered much later than they are introduced. 
Consequently, we can achieve that these methods struggle to identify new emerging vulnerabilities in real-world scenarios.

\begin{table*}[h]
\centering
\setlength{\tabcolsep}{0.6mm}
\renewcommand{\arraystretch}{1.1}
\caption{
The experimental results of vulnerability-related dependency prediction in the random and time-split settings. 
The shaded cells represent the performance of the best methods in each metric. Bold text cells represent the best performance. }
\resizebox{0.97\textwidth}{!}{
\begin{tabular}{cccccccccccccc}
\toprule
\rowcolor[HTML]{DEDEDE}
\multicolumn{2}{c}{\textbf{Split Methods}}    & \multicolumn{6}{c}{\textbf{Random}}                                                                       & \multicolumn{6}{c}{\textbf{Time}}                                                                         \\
\rowcolor[HTML]{DEDEDE}
\textbf{Type}             & \textbf{Baseline} & \textbf{Pre@1} & \textbf{Pre@3} & \textbf{Pre@5} & \textbf{Rec@1} & \textbf{Rec@3} & \textbf{Rec@5}\quad\quad & \textbf{Pre@1} & \textbf{Pre@3} & \textbf{Pre@5} & \textbf{Rec@1} & \textbf{Rec@3} & \textbf{Rec@5} \\
\toprule
Random                    & Random            & 56.90          & 68.60          & 75.26          & 30.48            & 55.46            & 68.67\quad\quad            & 55.36          & 68.11          & 78.67          & 32.26            & 62.42            & 75.77            \\
\midrule
\multirow{4}{*}{Lexical}  & Jaccard Similarity           & \cellcolor{gray!70}\textbf{69.70}          & \cellcolor{gray!20}70.68          & \cellcolor{gray!70}\textbf{77.94}          & \cellcolor{gray!70}\textbf{37.34}            & \cellcolor{gray!20}57.14            & \cellcolor{gray!70}\textbf{71.10}\quad\quad             & \cellcolor{gray!70}\textbf{68.04}          & \cellcolor{gray!20}72.33          & \cellcolor{gray!40}81.33          & \cellcolor{gray!70}\textbf{39.65}            & \cellcolor{gray!20}66.29            & \cellcolor{gray!40}78.33            \\
                          & Edit Similarity              & \cellcolor{gray!40}67.27          & \cellcolor{gray!70}\textbf{73.09}          & 76.51          & \cellcolor{gray!20}36.04            & \cellcolor{gray!70}\textbf{59.09}            & 69.81\quad\quad            & \cellcolor{gray!20}67.77          & \cellcolor{gray!70}\textbf{76.18}          & \cellcolor{gray!70}\textbf{86.17}          & \cellcolor{gray!40}39.49            & \cellcolor{gray!70}\textbf{69.82}            & \cellcolor{gray!70}\textbf{82.99}            \\
                          & BM25              & 62.42          & \cellcolor{gray!20}70.68          & \cellcolor{gray!20}76.87          & 33.44            & \cellcolor{gray!20}57.14            & \cellcolor{gray!20}70.13\quad\quad            & 63.91          & 71.45          & 80.50          & 37.24            & 65.49            & 77.53            \\
                          & BM25+          & \cellcolor{gray!20}67.27          & 70.28          & \cellcolor{gray!40}77.22          & \cellcolor{gray!20}36.04            & 56.82            & \cellcolor{gray!40}70.45\quad\quad            & 67.22          & 71.45          & 80.33          & 39.17            & 65.49            & 77.37            \\
                          \midrule
\multirow{2}{*}{Semantic} & CodeBERT          & \cellcolor{gray!40}68.48          & \cellcolor{gray!40}72.69          & 75.44          & \cellcolor{gray!40}36.69            & \cellcolor{gray!40}58.77            & 68.83\quad\quad           & 64.19          & \cellcolor{gray!20}72.33          & 78.00           & 37.4             & \cellcolor{gray!20}66.29            & 75.12            \\
                          & UnixCoder         & 61.82          & 69.48          & \cellcolor{gray!20}76.87          & 33.12            & 56.17            & \cellcolor{gray!20}70.13\quad\quad           & \cellcolor{gray!70}\textbf{68.04}          & \cellcolor{gray!40}73.03          & \cellcolor{gray!20}81.00          & \cellcolor{gray!70}\textbf{39.65}            & \cellcolor{gray!40}66.93            & \cellcolor{gray!20}78.01    \\    
                          \bottomrule
\end{tabular}}
\label{RQ2}
\end{table*}

\textbf{The performance of program analysis-
and prompt-based methods are not influenced under the
time-split setting.} 
Our empirical results indicate that program analysis achieved superior performance due to the predefined rules, with average F1 score and MCC from 7.85\%, 5.93\% in the random setting to 10.61\%, 8.69\% in the time-split setting, respectively. 
Besides, ChatGPT demonstrates near-optimal performance on both
F1 score and MCC metrics at
14.22\% and 10.13\%, respectively. Notably, ChatGPT is trained solely on data up to September 2021, which avoids the data leakage problem. 
The ChatGPT's performance can be attributed to its vast training corpus containing
general knowledge, which enables it to maintain consistent performance across diverse data distributions.

\begin{tcolorbox}[breakable,width=\linewidth-2pt,boxrule=0pt,top=1pt, bottom=1pt, left=4pt,right=4pt, colback=gray!15,colframe=gray!15]
\textbf{Summary for RQ1}: 
Experiment results reveal that fine-tuning-based methods exhibit superior performance in the random split setting. We also observe a performance degradation in supervised learning and fine-tuning-based baselines within a time-split setting. In addition, the program analysis and prompt-based methods are not influenced within the time-split setting, thereby preserving efficacy across real-world scenarios.
\end{tcolorbox}

\subsection{RQ2: Effectiveness in Vulnerability-related Dependency Prediction}
To answer RQ2, we 
evaluate
the performance 
of three types of methods
under both random-split and the time-split settings.
We present the top-1,3,5 Pre@k and Rec@k in Table~\ref{RQ2}.

\textbf{Superior performance of lexical-based methods for identifying dependency.}
The experimental results show that both lexical and semantic-based techniques enhances performance, yielding the averagely improvements of 10.21\% in Pre@1 and 5.74\% in Rec@1 for identifying dependency.
Notably, the lexical-based retrieval techniques contribute the largest improvements, with 
the consistent improvements 3.44\% $\sim$ 18.83\% and 3.45\% $\sim$ 18.91\% of Pre@k and Rec@k ($k=1,3,5$), respectively.
When
$k=1$, the performance benefits more from the knowledge associated with retrieval techniques. 
For instance, 
all retrieval methods show an average increase of 16.27\% in Pre@1, 3.72\% in Pre@3, and 2.06\% in Pre@5 compared to random method. 
Moreover, the semantic-based retrieval methods show moderate performance. This may be attributed to pre-trained models' focus on general semantics rather than domain-specific vulnerability knowledge,
indicating that incorporating vulnerability-specific characteristics in retrieval methods is beneficial.


\textbf{Jaccard Similarity and Edit Similarity outperform in random and time split settings, respectively.}
As 
shown in Table~\ref{RQ2}, 
Jaccard Similarity 
is the most effective method under
the random split setting, 
demonstrating superiority in 4 out of 6 cases. It achieves the best performance 69.70\% of Pre@1 and 37.34\% of Rec@1, respectively.
The Edit Similarity 
performs best 
in the time split setting, which outperforms the Jaccard Similarity by 3.85\% and 3.53\%, with respect to Pre@3 and Rec@3, respectively.
This finding implies that the retrieving common tokens between code snippets and vulnerability-related dependencies is effective on identifying dependency.

\begin{tcolorbox}[breakable,width=\linewidth-2pt,boxrule=0pt,top=2pt, bottom=2pt, left=4pt,right=4pt, colback=gray!15,colframe=gray!15]
\textbf{Summary for RQ2}: 
Our empirical analysis indicates that lexical-based methods yield superior results in identifying dependency. Specifically, the Jaccard Similarity and Edit Similarity achieve the best performance in the random and time-split settings, respectively.
\end{tcolorbox}

\begin{table*}[h]
\centering
\setlength{\tabcolsep}{2mm}
\renewcommand{\arraystretch}{1.03}
\caption{
The experimental results of repository-level vulnerability detection in the two split settings. 
The dark and light shaded cells represent the best performance by using vulnerability-related and predicted dependency, respectively. }
\resizebox{0.90\textwidth}{!}{
\begin{tabular}{cl@{\hspace{1\tabcolsep}}c@{\hspace{3\tabcolsep}}cccc@{\hspace{3\tabcolsep}}cccc}
\midrule
\rowcolor[HTML]{DEDEDE}
\multicolumn{3}{c}{\textbf{Split Methods}}                                                          & \multicolumn{4}{c}{\textbf{Random}}                                                       & \multicolumn{4}{c}{\textbf{Time}}                                                         \\
\rowcolor[HTML]{DEDEDE}
\textbf{Type}                & \multicolumn{1}{l}{\textbf{Baseline}}   & \multicolumn{1}{l}{\textbf{Strategy}}                                                  & \textbf{Precision}   & \textbf{Recall}      & \textbf{F1}    & \multicolumn{1}{l}{\textbf{MCC}}         & \textbf{Precision}   & \textbf{Recall}      & \textbf{F1}    & \textbf{MCC}         \\

\toprule
\multirow{11}{*}{Fine-tuning} & \multirow{2}{*}{CodeBERT}               & Upper     & 56.75              & 33.88           & 42.43             & 42.60        & 26.63              & 5.61            & 9.27              & 10.63        \\
                              &                                         & Prediction & 50.51              & 29.61           & 37.34             & 37.32        & 20.88              & 4.03            & 6.75              & 7.57         \\ \cline{2-11}
                              & \multirow{2}{*}{CodeT5}                 & Upper     & 52.82              & 37.76           & \cellcolor{gray!45}\textbf{44.04}             & \cellcolor{gray!45}\textbf{43.29}        & 23.59              & 9.75            & 13.79             & 12.93        \\
                              &                                         & Prediction & 49.66               & 32.14            & \cellcolor{gray!20}39.02              & \cellcolor{gray!20}38.54            & 17.03               & 5.72            & 8.56              & 7.73            \\ \cline{2-11}
                              & \multirow{2}{*}{UniXcoder}              & Upper     & 57.53              & 31.34           & 40.48             & 41.25        & 32.68              & 7.10            & 11.66             & 13.68        \\
                              &                                         & Prediction & 54.39              & 28.57           & 37.46             & 38.17        & 25.22              & 6.14            & 9.88              & \cellcolor{gray!20}10.72        \\ \cline{2-11}
                              & \multirow{2}{*}{PILOT}                   & Upper     & \cellcolor{gray!45}\textbf{69.78}              & 14.48           & 23.98             & 31.01        & 41.09              & 11.23           & 17.64             & 21.56        \\
                              &                                         & Prediction & \cellcolor{gray!20}68.00              & 12.65           & 21.33             & 28.57        & 20.00              & 2.33            & 4.17              & 5.57         \\ \cline{2-11}
                              & \multirow{2}{*}{LineVul}                & Upper     & 57.63              & 32.69           & 41.71             & 42.18        & 20.81              & 6.57            & 9.98              & 9.67         \\
                              &                                         & Prediction & 55.71              & 29.02           & 38.16             & 38.98        & 18.68              & 5.40            & 8.38              & 8.08         \\ \cline{2-11}
                              & \multirow{2}{*}{PDBERT}                 & Upper     & 57.72              & 34.03           & 42.82             & 43.09        & \cellcolor{gray!45}\textbf{47.08}              & 11.97           & \cellcolor{gray!45}\textbf{19.09}             & \cellcolor{gray!45}\textbf{22.26}        \\
                              &                                         & Prediction & 54.57              & 28.42           & 37.38             & 38.14        & \cellcolor{gray!20}26.43              & 3.92            & 6.83              & 8.82         \\
                              \hline{}
\multirow{12}{*}{Prompt}         & \multirow{2}{*}{LLaMA-7B}               & Upper     & 1.71               & 5.37            & 2.59              & -2.22        & 2.14               & 4.45            & 2.89              & -2.94        \\
                              &                                         & Prediction & 1.55               & 4.91            & 2.36              & -2.54        & 1.70               & 3.60            & 2.31              & -3.65        \\ \cline{2-11}
                              & \multirow{2}{*}{LLaMA-13B}              & Upper     & 2.04               & 15.22           & 3.60              & -2.65        & 2.32               & 11.12           & 3.84              & -4.32        \\
                              &                                         & Prediction & 2.06               & 15.33           & 3.63              & -2.60         & 2.35               & 11.12           & 3.88              & -4.21        \\ \cline{2-11}
                              & \multirow{2}{*}{CodeLlama-7B}       & Upper     & 2.35               & 29.70           & 4.36              & -2.41        & 2.67               & 23.83           & 4.80              & -5.31        \\
                              &                                         & Prediction & 2.11               & 26.79           & 3.91              & -3.56        & 2.65               & 23.20           & 4.76              & -5.29        \\ \cline{2-11}
                              & \multirow{2}{*}{CodeLLama-13B}      & Upper     & 2.19               & 27.16           & 4.05              & -3.09        & 2.83               & 24.05           & 5.06              & -4.51        \\
                              &                                         & Prediction & 2.05               & 24.85           & 3.79              & -3.69        & 2.94               & 24.79           & 5.25              & -4.10         \\ \cline{2-11}
                              & \multirow{2}{*}{GPT-3.5-instruct} & Upper     & 3.92               & \cellcolor{gray!45}\textbf{49.70}           & 7.27              & 4.70          & 5.18               & \cellcolor{gray!45}\textbf{56.46}           & 9.49              & 5.07         \\
                              &                                         & Prediction & 3.78 &
\cellcolor{gray!20}48.07& 7.00& 4.02& 5.03& \cellcolor{gray!20}42.27& 8.98& 3.53            \\ \cline{2-11}
                              & \multirow{2}{*}{ChatGPT}                & Upper     & 8.61               & 41.19           & 14.25             & 13.69        & 10.44              & 32.52           & 15.80             & 12.30         \\
                              &                                         & Prediction & 7.68               & 36.01           & 12.66             & 11.32        & 8.83               & 26.59           & \cellcolor{gray!20}13.26             & 9.03        \\
\bottomrule
\end{tabular}}
\label{RQ3}
\end{table*}
\subsection{RQ3: Effectiveness in Repository-level Vulnerability Detection}

This research question 
aims to investigate whether integrating
vulnerability-related dependencies can enhance the existing vulnerability detection methods' performance. 
We employ two strategies: ``Upper'' and ``Prediction'' to evaluate the baselines performance.
``Upper''  
refers to incorporate the vulnerability-related dependency as input for vulnerability detection. 
``Prediction''
represents the most effective retrieve method as identified in RQ2 (i.e., Jaccard Similarity in random split setting and Edit Similarity in time-split setting).
For the repository-level vulnerability detection, 
due to the input length limited, we only evaluate the fine-tuning- and prompt-based methods.
The experimental results are presented in Table~\ref{RQ3}.

\textbf{The incorporation of vulnerability-related dependency contexts improves vulnerability detection performance.} 
We observe that repository-level approaches that utilize the ``Upper'' strategy generally outperform function-level methods previously mentioned. Specifically, when applying the ``Upper'' strategy in fine-tuning-based methods, we can observe performance enhancements in five out of six baselines in \tool. Except for PILOT, these repository-level methods demonstrate an average improvement over the corresponding baselines of 7.43\% in precision, 3.38\% in recall, 4.91\% in F1 score, and 5.24\% in MCC. This suggests that incorporating vulnerability-related dependencies provides additional contextual information, which allows the model to leverage a more comprehensive understanding of the code repository. The observed performance decline in PILOT may be attributed to its weakly supervised learning, which can be a consequence of an excess of unlabeled samples in the dataset.

\textbf{Larger models benefit more from the repository-level vulnerability-related knowledge.} Our experimental findings indicate that the benefits derived from repository-level information are marginal for the LLaMA and CodeLlama. It may be due to the
limitations of the model's abilities for capturing vulnerability patterns. In contrast, ChatGPT exhibits performance improvements across all four evaluation metrics in combing repository-level dependencies, with increases of 11.60\%, 24.43\%, 14.48\%, and 26.35\%, respectively. These results suggest that models with a larger foundational architecture possess superior comprehension abilities when dealing with extensive textual input.

\textbf{It is imperative to explore more effective retrieval methods for identifying dependency.}
Despite utilizing the most effective retrieval approach identified in RQ2 for identifying dependencies, combining it with existing repository-level vulnerability detection techniques does not greatly
enhance the performance. For instance, under a random setting, ChatGPT employing Edit Similarity yields
improvements of 4.07\%, 10.63\%, 5.24\%, and 8.43\% respectively across the four metrics. However,  using Edit Similarity under the time-split setting is not effective. These observations underscore the need for advancements in retrieval strategies
to better capture and leverage vulnerability-related dependency.

\begin{tcolorbox}[breakable,width=\linewidth-2pt,boxrule=0pt,top=2pt, bottom=2pt, left=4pt,right=4pt, colback=gray!15,colframe=gray!15]
\textbf{Summary for RQ3}: 
The experiment results reveal that incorporating contexts related to vulnerabilities enhances the performance of vulnerability detection. It is noteworthy that larger models particularly gain improvement from the integration of vulnerability knowledge at the repository level. In addition, it becomes essential to develop more effective retrieval techniques for identifying vulnerability-related dependencies.
\end{tcolorbox}

\subsection{RQ4: Effectiveness in Each CWE Type Vulnerability Detection}

\begin{figure*}
    
  \begin{minipage}{0.195\textwidth}
    \centering
    \includegraphics[width=\textwidth]{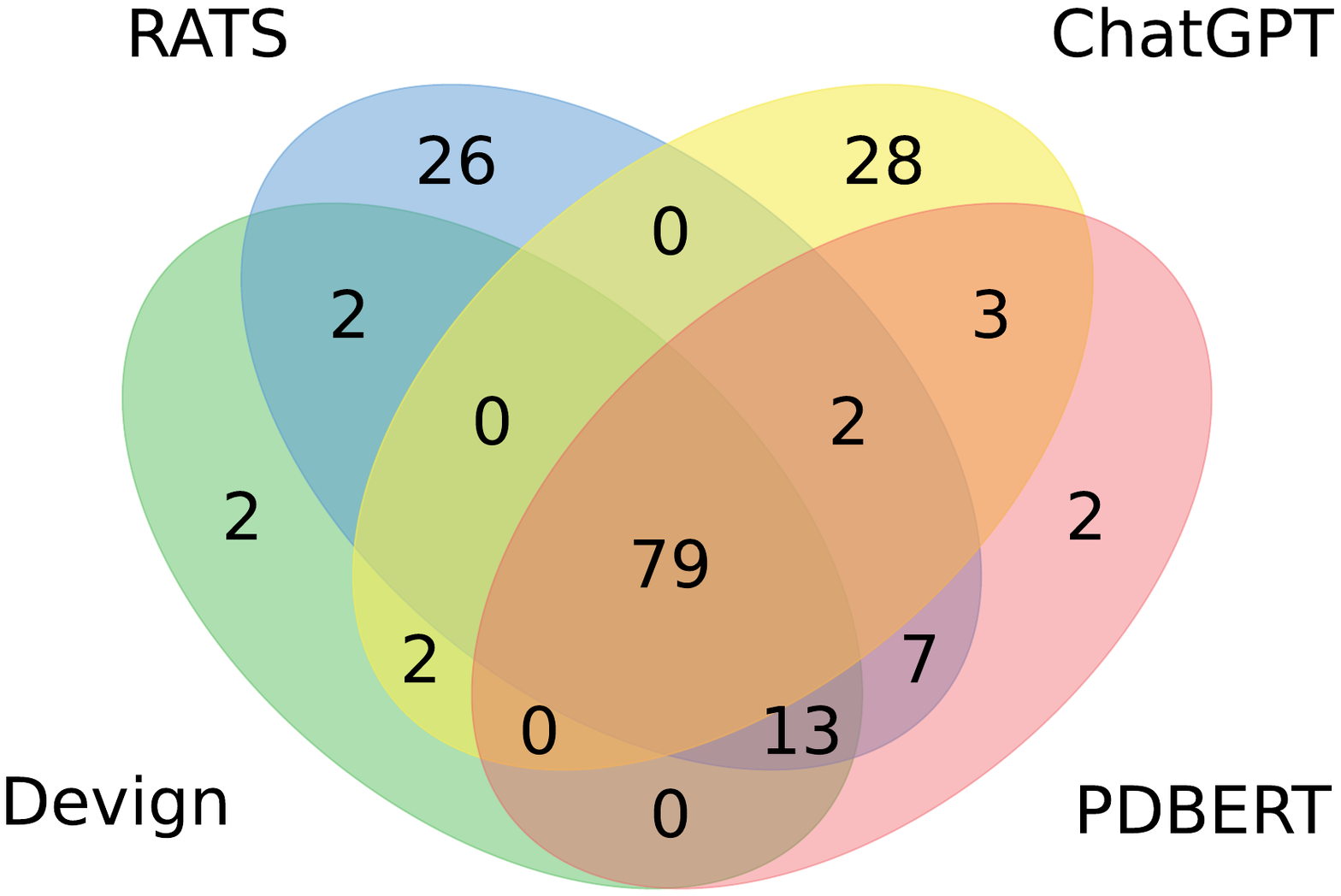}
    \subcaption{CWE-190}
  \end{minipage}
  \begin{minipage}{0.195\textwidth}
    \centering
    \includegraphics[width=\textwidth]{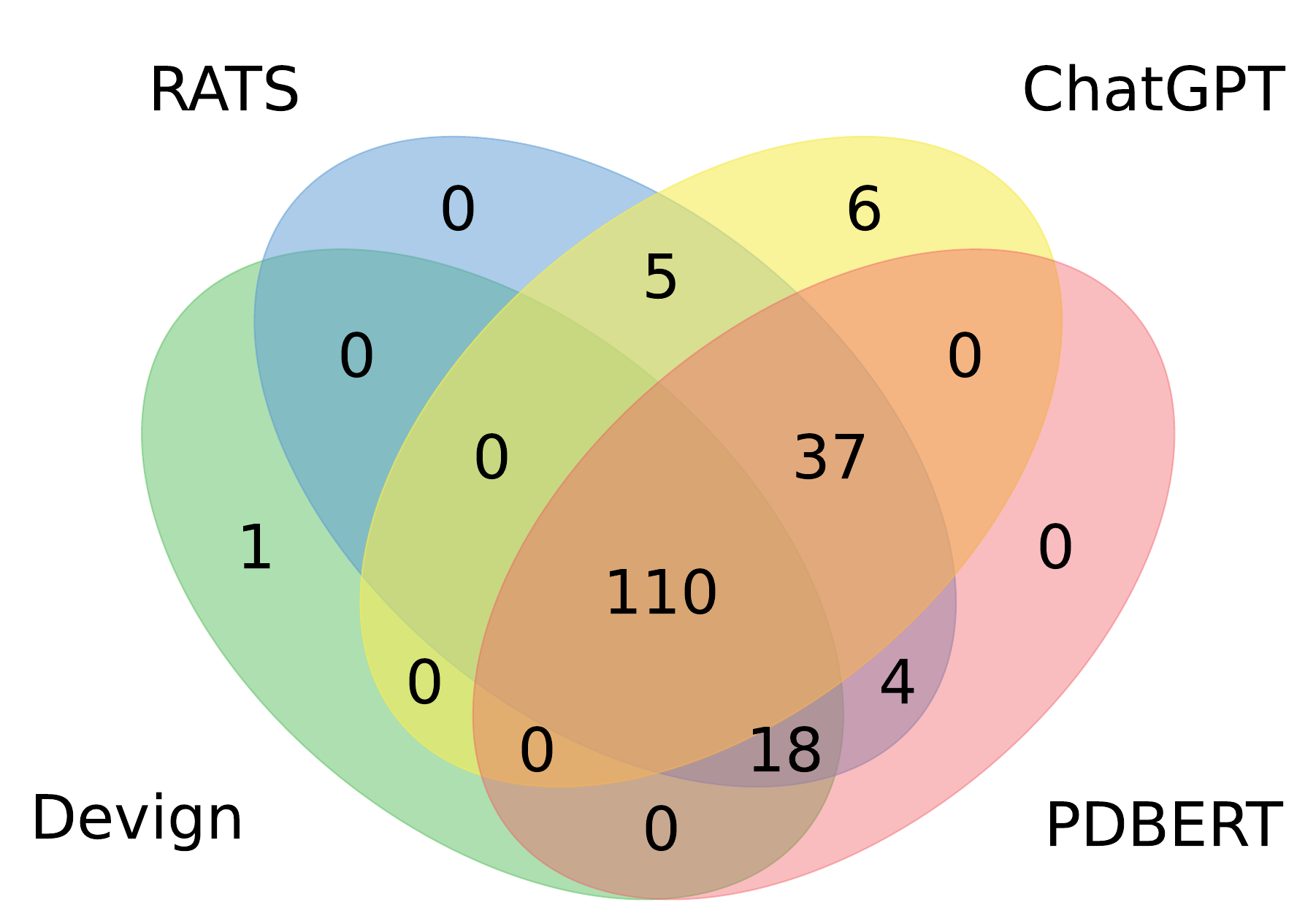}
    \subcaption{CWE-400}
  \end{minipage}
  \begin{minipage}{0.195\textwidth}
    \centering
    \includegraphics[width=\textwidth]{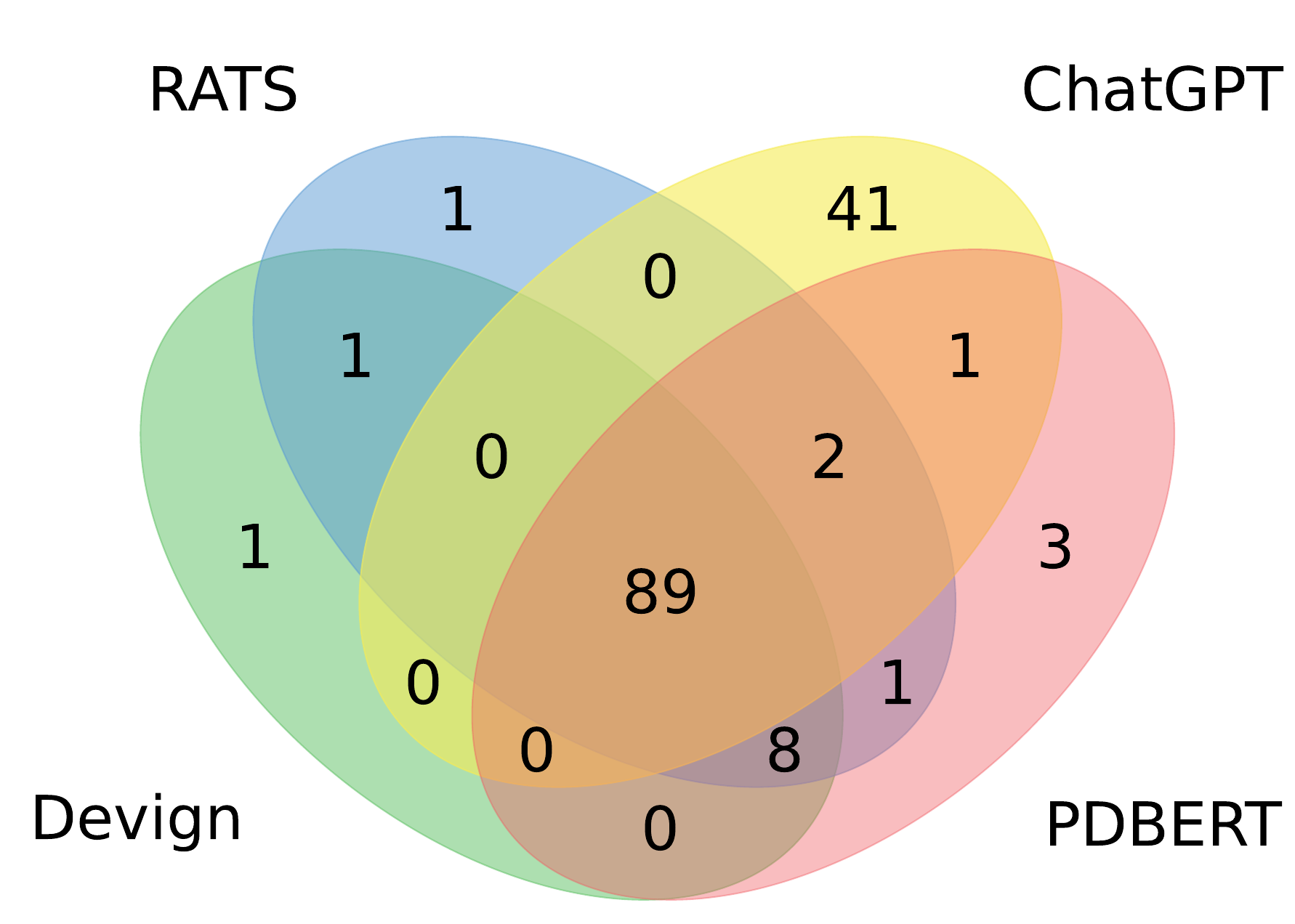}
    \subcaption{CWE-415}
  \end{minipage}
  \begin{minipage}{0.195\textwidth}
    \centering
    \includegraphics[width=\textwidth]{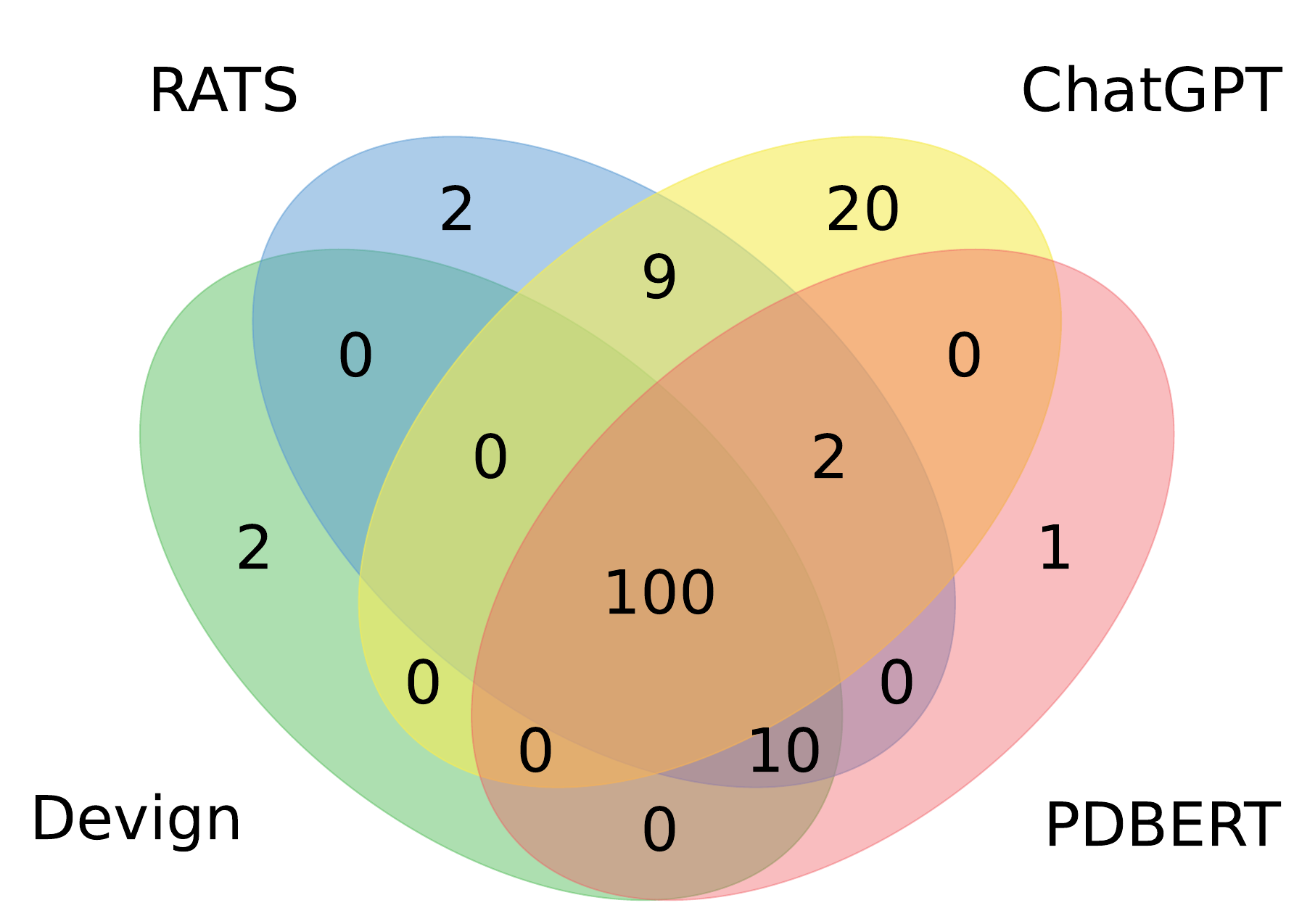}
    \subcaption{CWE-416}
  \end{minipage}
  \begin{minipage}{0.195\textwidth}
    \centering
    \includegraphics[width=\textwidth]{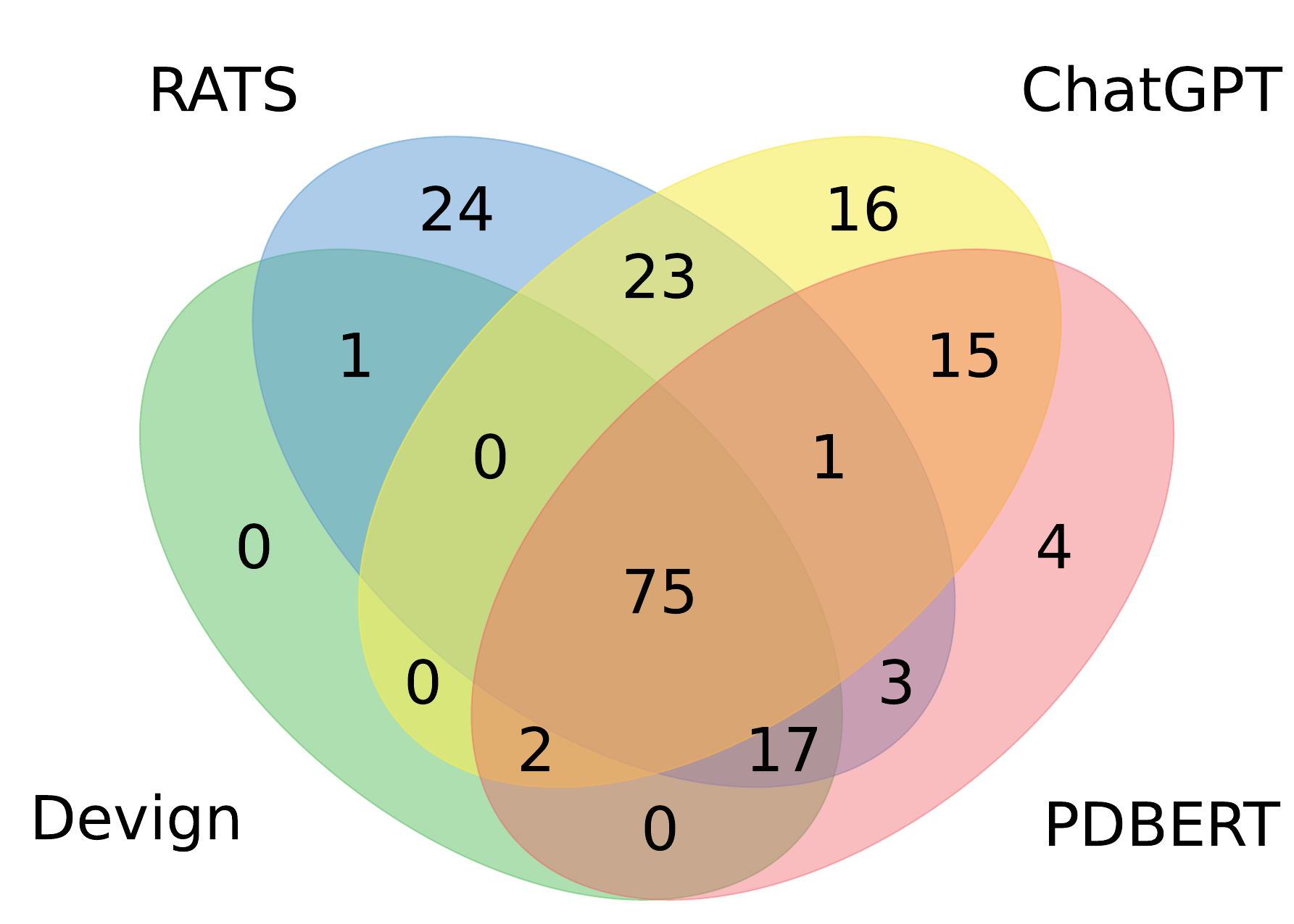}
    \subcaption{CWE-787}
  \end{minipage}
  \caption{The experimental results of several vulnerability types, including CWE-190, CWE-400, CWE-415, CWE-416, and CWE-787. The green, blue, red, and yellow circles denote the results of Devign, RATS, PDBERT, and ChatGPT, respectively.}
  \label{RQ4}
\end{figure*}

To answer RQ4, we select four methods from different types of methods (i.e., RATS, Devign, PDBERT, and ChatGPT), which perform the best overall performance in their types. 
We then evaluate these methods in CWE-190, CWE-400, CWE-415, CWE-416, and CWE-787.
These vulnerability types represent the most recurrent vulnerabilities, highlighting their elevated potential for software damage. 
For each type, we deliberately choose 200 representative samples within the time-split setting to avoid the problems of data leakage.
Figure~\ref{RQ4} shows the number of correctly predicted samples by the four baselines on each of the vulnerability type.

\textbf{The superior performance of ChatGPT for each CWE vulnerability detection.}
In the domain of singular vulnerability detection, our analysis reveals that ChatGPT has demonstrated superior performance, correctly identifying 668 samples and achieving 47.53\% F1 score averagely. For example, within the CWE-416, ChatGPT can correctly detect 131 samples and achieve 45.67\% F1 score respectively, demonstrating effectiveness superior to general vulnerability detection. Moreover, ChatGPT exclusively identified 20 samples, while only 2, 2, and 1 samples can be detected by RATS, Devign, and PDBERT respectively. Therefore, it is practical to leverage LLMs to design a detector for specific CWE type vulnerability in real-world scenarios. 

\textbf{
It is worthwhile to explore how to combine the vulnerability detection capabilities of
different baselines.
} Among the 200 samples in CWE-787, the four baselines correctly detect 122 samples in average. The capabilities of the four models are evidenced by their ability to correctly predict 181 samples, showcasing their complementary strengths. Concurrently, there exists a subset of 75 samples that are detectable by any of the four baselines, illustrating the overlap in their detection capabilities. In the future, 
it is worthwhile to explore how to combine the vulnerability detection capabilities of different methods.

\begin{tcolorbox}[breakable,width=\linewidth-2pt,boxrule=0pt,top=0pt, bottom=0pt, left=4pt,right=4pt, colback=gray!15,colframe=gray!15]
\textbf{Summary for RQ4}: 
The experimental results reveal that the superior performance of ChatGPT for each CWE type vulnerability detection.
Besides, it is worthwhile to explore how to combine different methods' capabilities for software vulnerability detection in the future.
\end{tcolorbox}

\section{Discussion}
\label{sec:discussion}

\subsection{Implications of Findings}
In this section, we discuss the implications of our work for software vulnerability detection.
Our experimental results also show potential research directions in the era of software vulnerability detection.
Specifically:

\begin{enumerate}
    \item For RQ1, fine-tuning-based methods exhibit superior performance in the random split setting. They need to consider time factors to be more effective in real-world scenarios. The program-analysis- and prompt-based methods effectively are not affected by time-split setting.
    Leveraging LLMs and prompt techniques can be a solution for alleviating performance degradation by time-split setting, thereby enhancing the applicability in real-world scenarios.
    \item For RQ2, using lexical-based methods for identifying vulnerability-related dependency leads to relatively better performance than other semantic-based methods. However, the number of dependencies is not consistent across the code samples. Therefore, how to automatically identify the dependencies needs to be further investigated.
    \item  For RQ3, incorporating contexts related to vulnerabilities in repository-level vulnerability detection enhances the performance compared with function-level vulnerability detection. 
    Moreover, the larger LLMs benefit more from the repository-level vulnerability-related knowledge.
    The retrieval techniques for predicting vulnerability-related dependency are one major bottleneck for improving the performance of current repository-level approaches and remain unsolved.
    \item  For RQ4, ChatGPT is more effective than other vulnerability detection methods for detecting the specific CWE vulnerability types. 
    In addition, it is worthwhile to explore how to combine different methods’ capabilities for software vulnerability detection in the future.


\end{enumerate}

\subsection{Threats to Validity}
\label{validity}
\textit{Representativeness of Baselines Selection.} 
A potential threat to the validity of our study arises from the representativeness of the baselines employed in our experiments for vulnerability detection.
Owing to the constraints posed by computational resources and the excessively expensive costs associated with the usage of APIs, we refrained from conducting experiments involving a 34B model size, as well as several other contemporary models including CodeGeex~\cite{CodeGeeX}, StarCoder~\cite{DBLP:journals/corr/abs-2305-06161/starcoder}, and GPT-4~\cite{GPT4}. Future research will conduct more comprehensive experiments across broader baselines.

\textit{Generalizability on Other Programming Languages.} 
In this paper, 
our experimental analysis focuses solely on C/C++ programming languages, excluding other popular languages such as Java and Python. However, the system of \tool can be generalized to other programming languages because the approach does not rely on language-specific features. In future research, we intend to evaluate the efficacy of \tool in the context of a broader range of programming languages.

\textit{Implementation of baselines.}
To replicate the baselines, we meticulously use the methodologies delineated in the open-source codes and the original papers. However, owing to the unavailability of the implementation details and hyper-parameters for Devign~\cite{devign}, our reproduction is guided by Reveal's implementation~\cite{reveal}.

\section{Related Work}
\label{sec:related}
We have elaborated on the vulnerability detection methods in Section~\ref{sec:background}, and focus on illustrating the vulnerability datasets in this section.
These datasets can be broadly classified into three groups:
function-, slice- and file-level.
The function-level datasets~\cite{devign, DBLP:conf/msr/FanL0N20/fan, diversevul}  utilize cases crafted artificially or extract the functional source code from the real-world scenarios code snippets. For example, SARD~\cite{SARD} constructs samples by manual checking and industrial production.  Reveal~\cite{reveal} collects patches from open-source repositories and extracts function-level data from code changes in the patches. The primary limitation of these function-level datasets is constrained by the context-provided code segments.
The slice-level datasets~\cite{deepbugs, DeepWukong}  typically employ pre-defined rules or static tools to extract DFG and CFG from the source code.  For example, Vuldeepecker~\cite{DBLP:conf/ndss/LiZXO0WDZ18/vuldeepecker} constructs Code Gadget Database (CGD) by generating code gadgets, consisting of CWE-119 and CWE-399 vulnerabilities. $\mu$Vuldeepecker~\cite{uVulDeePecker} expands upon this approach by collecting the 40 types of vulnerabilities and their corresponding labels. 
The file-level datasets~\cite{file1, file2, file3} generally provide entire files from vulnerability patches. Such datasets present a comprehensive snapshot of the source code, which is beneficial in detecting the broader context within which vulnerabilities may exist. For example, CrossVul~\cite{CrossVul} constructs the dataset spanning 40 programming languages and 1,675 projects, while it only provides file-level source code. 

However, the previous works more focus on collecting function/file-level data, and ignore repository-level dependencies which are important for detecting inter-procedural vulnerability.

\section{Conclusion and Future Work}
\label{sec:conclusion}

In this paper, we propose a holistic multi-level evaluation system \tool, aiming at evaluating the software vulnerability detection performance of inter- and intra-procedural vulnerabilities simultaneously. 
Specifically, \tool consists of three evaluation tasks: 
function-level vulnerability detection, 
vulnerability-related dependency prediction
and repository-level vulnerability detection.
\tool also consists of a large-scale vulnerability dataset.
By evaluating 19 vulnerability detection methods on the data split randomly and by time respectively, we observe that incorporating vulnerability-related dependencies facilitates repository-level vulnerability detection performance compared with function-level vulnerability detection. Our analysis highlights the current progress and future directions for software vulnerability detection. 
In the future, we will explore more aspects of repository-level vulnerability detection such as designing retrieval methods for identifying vulnerability-related dependencies and integrating dependency information in prompts.


\balance
\bibliographystyle{ACM-Reference-Format}
\bibliography{Citation}









\end{sloppypar}

\end{document}